\documentclass[10pt,twocolumn,prl,english]{revtex4}
\usepackage[T1]{fontenc}
\usepackage[latin9]{inputenc}
\usepackage{float}
\usepackage{graphicx}
\usepackage{amssymb}
\usepackage{amsmath}
\RequirePackage{xspace}

\makeatletter

\usepackage{color}
\usepackage{graphicx}

\newcommand{\BA}{{\bf A}}
\newcommand{\BS}{{\bf S}}
\newcommand{\BY}{{\bf Y}}
\newcommand{\BX}{{\bf X}}
\newcommand{\Bf}{{\bf f}}
\newcommand{\Bg}{{\bf g}}
\newcommand{\ie}{i.\,e.\@\xspace}
\newcommand{\eg}{e.\,g.\@\xspace}
\newcommand{\vs}{vs.\@\xspace}
\newcommand{\Fig}{Fig.\@\xspace}
\newcommand{\Eq}{Eq.\@\xspace}
\newcommand{\Eqs}{Eqs.\@\xspace}

\newcommand{\shalf}{{\textstyle \frac{1}{2}}}

\usepackage{babel}
\makeatother

\begin{document}

\title{Bubbling in delay-coupled lasers}

\author{V. Flunkert\textsuperscript{1}}
\author{O. D'Huys\textsuperscript{2}}
\author{J. Danckaert\textsuperscript{2,3}}
\author{I. Fischer\textsuperscript{4}} 
\author{E. Sch\"oll\textsuperscript{1}} 

\affiliation{\textsuperscript{1}Institut f{\"u}r Theoretische Physik, 
TU Berlin, Hardenbergstra\ss{}e 36, 10623 Berlin, Germany}

\affiliation{\textsuperscript{2}Department of Physics (DNTK), 
Vrije Universiteit Brussel, Pleinlaan 2, 1050 Brussel, Belgium}

\affiliation{\textsuperscript{3}Dept.~of Appl.~Phys.~and Photon.~(TONA), Vrije Universiteit Brussel, Pleinlaan 2, 1050 Brussel, Belgium}

\affiliation{\textsuperscript{4}School of Engineering and Physical Sciences, 
Heriot-Watt University, Edinburgh EH1 4AS, Scotland, UK}


\begin{abstract}
  We theoretically study chaos synchronization of two lasers
  which are delay-coupled via an active or a passive relay. While the lasers
  are synchronized, their dynamics is identical to a single laser with delayed
  feedback for a passive relay and identical to two delay-coupled lasers for
  an active relay. Depending on the coupling parameters the system exhibits
  bubbling, \ie, noise-induced desynchronization, or on-off intermittency.  We
  associate the desynchronization dynamics in the coherence collapse and low
  frequency fluctuation regimes with the transverse instability of some of the
  compound cavity's antimodes. Finally, we
  demonstrate how, by using an active relay, bubbling can be suppressed.
\end{abstract}

\maketitle

Synchronization phenomena of coupled nonlinear oscillators are omnipresent and
play an important role in physical, chemical and biological systems
\cite{BOC02,PIK01}.  Understanding the synchronization mechanisms is 
crucial for many practical applications.  One of the most interesting and
challenging phenomena when coupling nonlinear systems is the synchronization of
chaotic dynamics \cite{PEC90}. 
In order to characterize the synchronization effects, stability properties are
a key issue. Noise can, for instance, cause intermittent desynchronization.
This behavior is called bubbling \cite{ASH94} and has been observed for
example in optical \cite{TER99,SAU98} and electrical \cite{GAU96} systems.

Semiconductor lasers are of particular interest in the study of chaos
synchronization. The synchronization properties may facilitate new secure
communication schemes. However, if two identical semiconductor lasers are
optically coupled over a finite distance, it has been observed that the
coupling delay leads to spontaneous symmetry breaking, and only generalized
synchronization of leader-laggard type occurs \cite{MUL04}. A passive relay in
form of a semitransparent mirror or an active relay in form of a third laser in
between the two lasers have been shown to stabilize the isochronous
synchronization solution \cite{SHA06,KLE06,LAN07,FIS06}, rendering such
configurations attractive for chaos based applications, like, \eg,
bidirectional encrypted communication, or chaos-based key exchange, as detailed
in ref.~\cite{VIC04}.

In this work we show theoretically that bubbling and on-off intermittency occur in both
relay setups. In the coherence collapse (CC) and in the low frequency
fluctuation (LFF) regime, we find that bubbling is caused by transversally
unstable external cavity modes (ECMs). In the LFF regime the localization of the
transversally unstable modes in the synchronization manifold (SM) results in
desynchronization during power dropouts, which has also been observed in
unidirectionally coupled lasers \cite{AHL98}. For the active relay we find that
bubbling can be suppressed by stronger pumping of the relay laser.

\begin{figure}[h]
  \centering
  \includegraphics%
  [width=1.0\columnwidth]%
  {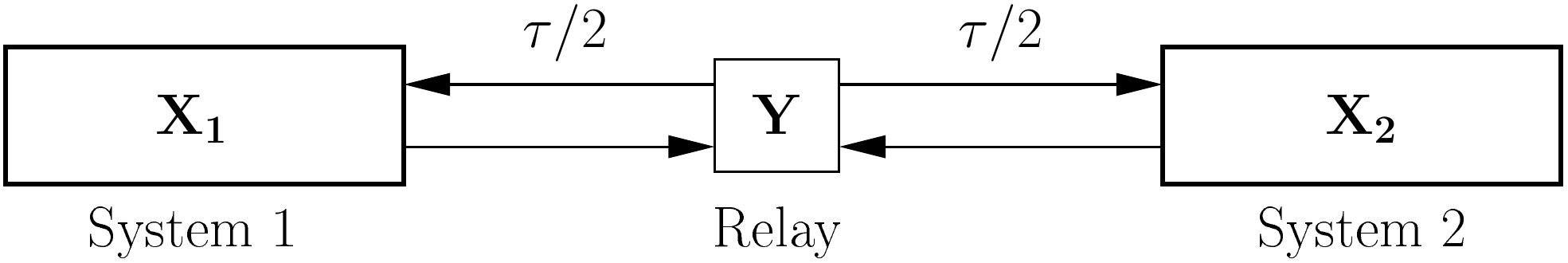}
  \caption{Schematic setup.}
  \label{fig:setup}
\end{figure}

We consider two identical systems which are delay-coupled 
via a relay (\Fig\ref{fig:setup}).
The relay may be an active element or a passive element which merely
distributes the arriving signals between the systems.  
Each system receives a delayed signal from the relay
\begin{align}
  \dot{\BX}_{j} & = \Bf(\BX_{j})+K\,\BY(t-\tau/2)\qquad ({\textstyle j = 1, 2}). \label{eq:general}
\end{align}
Here $\BX_j, \BY\in \mathbb{R}^n$ are the state vectors of the system $j$ and
the relay, respectively, $\Bf$ is a nonlinear function, $K$ is the
relay-to-system coupling matrix, and $\tau$ is the propagation delay between
system 1 and system 2. The overdot denotes the derivative with respect to time $t$. 

For the active relay we consider the equation
\begin{align}
  \dot{\BY} & = \Bg(\BY) + \shalf L\, \BX_{1}(t-\tau/2) + \shalf L\, \BX_{2}(t-\tau/2), \label{eq:active}
\end{align}
where $L$ is the system-to-relay coupling matrix and the function $\Bg$
describes the internal dynamics of the relay. For the passive relay we consider the
algebraic equation
\begin{align}
  \BY(t) & = \shalf [ \BX_{1}(t-\tau/2) + \BX_{2}(t-\tau/2) ]. \label{eq:passive}
\end{align}
Equation (\ref{eq:general}) together with the relay equation
(\ref{eq:active}) or (\ref{eq:passive}) allow for an
isochronous (or \textit{zero-lag}) solution $\BX_1(t)=\BX_2(t)$, respectively.  The SM
is thus invariant. To analyse the stability of this solution we introduce a
symmetric variable $\BS = \shalf(\BX_{1}+\BX_{2})$ and an antisymmetric
variable $\BA = \shalf(\BX_{1}-\BX_{2})$.
Equation (\ref{eq:general}) can then be rewritten in the new variables
  \begin{align}
    \dot{\BS}&= \shalf \left[\Bf(\BS+\BA)+\Bf(\BS-\BA)\right]+ K\, \BY(t-\tau/2), \label{eq:fullS}\\
    \dot{\BA}&= \shalf \left[\Bf(\BS+\BA)-\Bf(\BS-\BA)\right]. \label{eq:fullA}
  \end{align}
Note that due to the symmetric coupling the delay terms and all the coupling
parameters in \Eq(\ref{eq:fullA}) vanish. Equation (\ref{eq:fullA}) taken at
$\dot\BA={\bf 0}$ has a solution $\BA={\bf 0}$ which represents the
isochronously synchronized state. Its stability is determined by
linearizing \Eqs (\ref{eq:fullS}) and (\ref{eq:fullA}) in the variable $\BA$ around $\BA={\bf
0}$, \ie, we linearize orthogonal to the SM:
  \begin{align}
    \dot{\BS}&= \Bf(\BS)+ K\, \BY(t-\tau/2), \label{eq:linearS}\\
    \dot{\BA}&= D\Bf(\BS)\BA. \label{eq:linearA}
  \end{align}
Here, $D\Bf(\BS)$ denotes the Jacobian of $\Bf$ evaluated at the position
$\BS$. Since $\BS$ depends on time, \Eq(\ref{eq:linearA}) constitutes a
time-dependent variational equation.

For both relay types the dynamics within the SM resembles the dynamics of a
single system with either self-feedback (passive relay) 
\begin{align} 
  \dot \BS &= \Bf(\BS) + K \BS(t-\tau)\label{eq:DynPassiveRelay} 
\end{align} 
or coupling to the active relay 
\begin{align} 
  \dot \BS &= \Bf(\BS) + K\, \BY(t-\tau/2),\label{eq:DynActiveRelay}\\ 
  \dot \BY &= \Bg(\BY) + L\, \BS(t-\tau/2).
\end{align}
In both cases the stability of the synchronized solution is governed by
\Eq(\ref{eq:linearA}). However, the trajectory $\BS(t)$ will be different and
the synchronized state may thus have different stability properties.

Bubbling occurs \cite{ASH94,VEN96a} when an invariant set $I$, for example a
periodic orbit, in the SM is transversally unstable, while the chaotic
attractor in the SM is still transversally stable, \ie the largest transversal
Lyapunov exponent of the attractor is negative, $\lambda_{\perp}<0$. In this
situation the trajectory can be pushed towards the unstable set by noise and
leave the SM. If there is no other attractor present, the trajectory will
eventually come back to the SM and the systems will synchronize again.  The point
where the invariant set $I$ loses its transverse stability is called
bubbling bifurcation, while the point where the attractor itself becomes
unstable is called blow-out bifurcation.

For semiconductor lasers the dynamics of each system is governed by the
dimensionless Lang-Kobayashi rate equations \cite{LAN80b,ALS96}
\begin{align}
  \dot{E_j} & = \shalf(1+i\alpha)n_j\, E_j + K e^{i\varphi} E_{\BY}(t-\tau/2) + F_j(t)\nonumber \\
  T \dot{n_j} & = p-n_j-(1+n_j)\,|E_j|^{2}. \label{eq:laser}
\end{align}
Here, $E_j$ and $E_\BY$ are the complex electric field amplitudes of the $j$th
system and the relay, respectively, $n_j$ is the excess carrier density,
$\alpha$ is the linewidth enhancement factor, $p$ is the pump current, and the
timescale parameter $T=\tau_c/\tau_p$ is the ratio of the carrier ($\tau_c$)
and the photon $(\tau_p)$ lifetime. For simplicity we choose the feedback phase
$\varphi=0$. Note that in general one could also include coupling phases in
\Eq(\ref{eq:passive}). This leads to interference conditions of all phases
which have to be satisfied for isochronous synchronization.  In our simulations we consider
the spontaneous emission noise via a complex Gaussian white random variable
$F_j(t)$ with the covariance $\langle F_j(t)\, F_i(t')^{*}\rangle=\beta (n+n_0) \delta_{ij}
\delta(t-t')$, where $n_0=10$ is the carrier density at threshold and
$\beta=10^{-5}$ is the spontaneous emission factor.  Carrier noise has not been
taken into account at this level.

\begin{figure}[h]
  \centering
  \includegraphics%
  [width=0.9\columnwidth]%
  {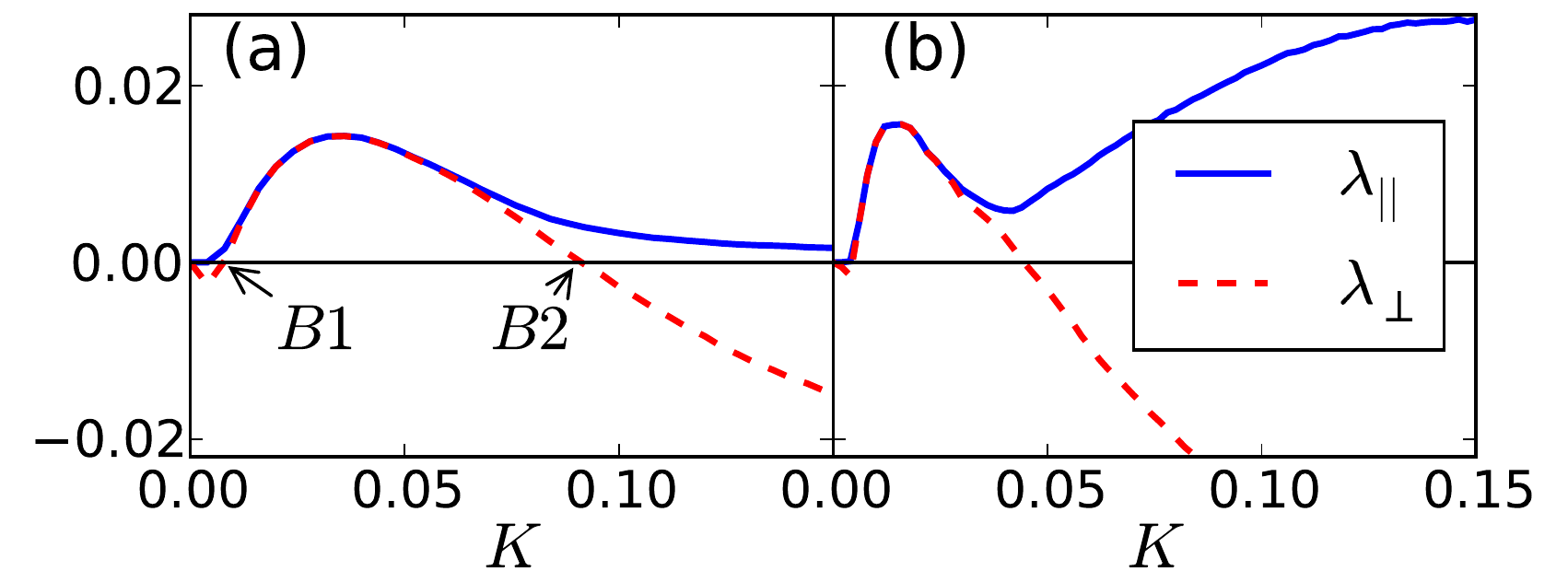}
  \caption{(Color online) Maximum transversal Lyapunov exponent $\lambda_\perp$
  (red dashed) and maximum parallel Lyapunov exponent $\lambda_{||}$ (blue
  solid) as a function of the feedback strength $K$ for a) passive relay b) active relay
  ($p_{\mbox{\scriptsize relay}}=4.0$).
  At the two blow-out bifurcations $B1$ and $B2$ the maximum transversal
  Lyapunov exponent of the chaotic attractor changes sign.
  Other parameters: $T=200$, $p=1.0$, $\tau=1000$, $\alpha=4$, $\varphi=0$}
  \label{fig:lyap}
\end{figure}

\begin{figure}[h] \centering
  \includegraphics%
  [width=\columnwidth]%
  {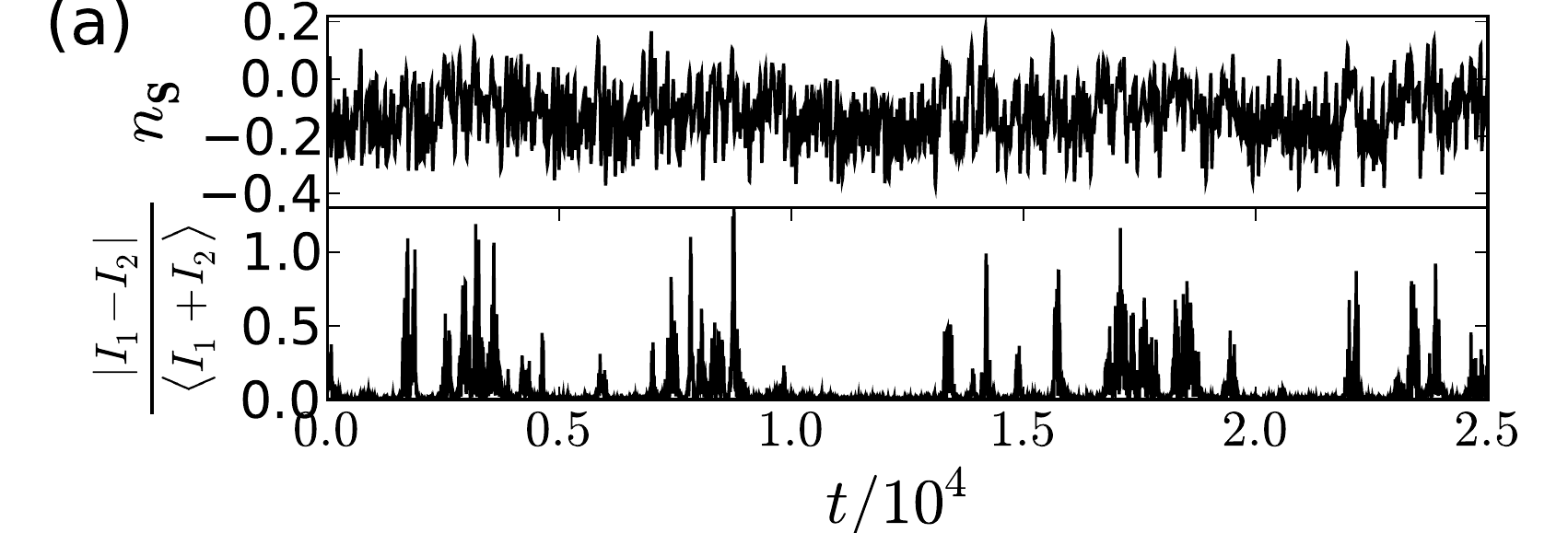} 
  \includegraphics%
  [width=\columnwidth]%
  {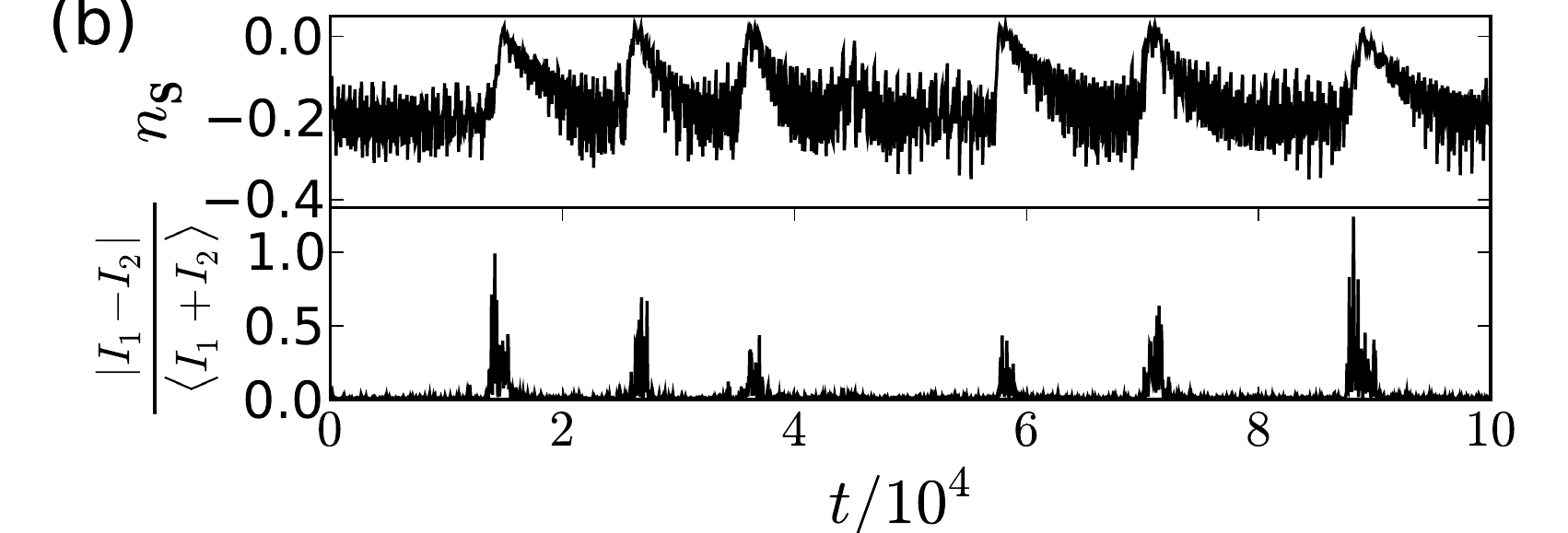} 
  \caption{
  Carrier density of the symmetric variable $n_\BS=\shalf(n_1 +n_2)$ and
  intensity difference $|I_1-I_2|/\langle I_1+I_2\rangle$ (normalized by the
  mean intensity) representing the deviation from the synchronized state \vs
  time. 
  a): Bubbling in the coherence collapse regime ($p=1.0$).
  b): Bubbling in the low frequency fluctuation regime during 
  power dropouts ($p=0.1$).
  Other parameters: $T=200$, $K=0.12$, $\tau=1000$, $\alpha=4$.\\ } 
  \label{fig:bubble}
\end{figure}

If the relay is realized through a semitransparent mirror (passive relay), the
dynamics within the SM is given by \Eqs(\ref{eq:laser}) with
$E_\BY(t-\tau/2)=E_j(t-\tau)$, \ie, an effectively decoupled laser. For this
configuration we calculate the maximum parallel Lyapunov exponents
$\lambda_{||}$ (within the SM) as well as the maximum transversal Lyapunov
exponents $\lambda_\perp$ by simulating the dynamics in the SM without noise
and applying the method developed in \cite{FAR82}. Figure \ref{fig:lyap}a
displays the Lyapunov exponents as a function of the feedback strength $K$.
There are two blow-out bifurcations \cite{OTT94} at $K\approx 0.008$ ($B1$) and
at $K \approx 0.09$ ($B2$), where $\lambda_\perp$ changes sign and the chaotic
attractor loses its transversal stability. Similar behavior is found for an
active relay (\Fig\ref{fig:lyap}b).
Over a wide range of $K$ (\Fig\ref{fig:lyap}a) in which the attractor is stable
and the dynamics is chaotic, we observe bubbling induced by spontaneous
emission noise. In these regimes, when the noise is switched off in the
simulations, the two lasers stay perfectly synchronized. In the regime with
$\lambda_\perp>0$ we observe desynchronization bursts even without noise, \ie,
the system exhibits on-off intermittency.  Figure \ref{fig:bubble}a depicts the
bubbling behavior for values of $K$ above $B2$ where the laser operates in the
CC regime. Figure \ref{fig:bubble}b corresponds to a lower pump current, where
the synchronized lasers operate in the LFF regime. In this regime bubbling only
takes place during the power dropouts. In both cases, when the noise amplitude
is decreased, the desynchronization peaks occur less frequently, the maximum
height, however, does not decrease.

\begin{figure}[h]
  \centering
  \includegraphics%
  [width=1.0\columnwidth,keepaspectratio]%
  {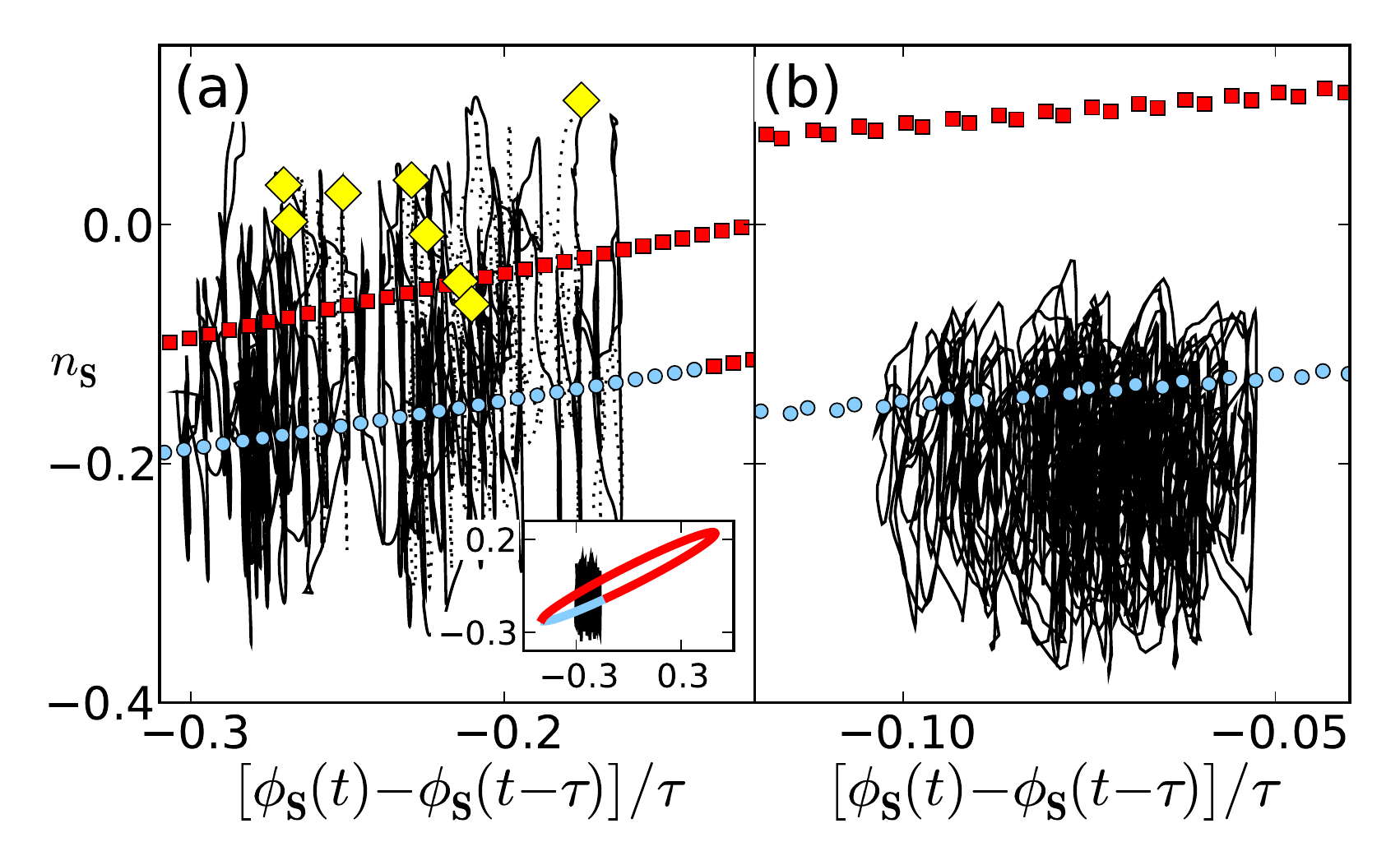} 
  \caption{(color online) Projection of the dynamics of the symmetrized
  solution $n_\BS$, $E_\BS=A_\BS\exp(i\phi_\BS)$ (black trajectory) onto the
  $(\omega, n)$-plane for a) passive relay ($p=1.0$), b) active relay ($p=1.0$,
  $p_{\mbox{\scriptsize relay}}=4.0$). Transversally stable (blue circles) and
  transversally unstable (red squares) ECMs are also shown.  a) The competition
  between chaotic itinerancy and antimodes leads to bubbling during global
  antimode dynamics. Yellow diamonds mark the onset of desynchronization.
  Solid and dashed parts of the trajectory correspond to synchronized and
  desynchronized periods, respectively.  The inset in a) shows the ECM ellipse
  and bubbling dynamics in a larger range. b) The system evolves around the
  transversally stable compound laser modes and bubbling is suppressed.
  Parameters as in \Fig\ref{fig:bubble}.
  \label{fig:ecmdynamics}}
\end{figure}

We now relate the desynchronization dynamics to the transverse stability of the
ECMs in the SM. These modes organize the dynamics in the SM in the CC and the
LFF regime. The ECMs are rotating wave solutions of the form $E(t)=A
\exp(i\omega t)$ and $n(t)=n$ with constant values $A$, $\omega$ and $n$. They
are well studied \cite{MOR92} solutions of the Lang-Kobayashi equations and are
located on an ellipse in the $(\omega, n)$-plane (see inset of
\Fig\ref{fig:ecmdynamics}a).  The modes on the top and bottom half of the
ellipse are called modes and antimodes, respectively. 

The transverse stability of an ECM is governed by the variational equation
(\ref{eq:linearA}) where $\BS(t)$ is the ECM solution.  To determine the
stability, we transform the laser equations into a rotating frame \cite{YAN05}
$E_0=E \exp(-i\omega t)$. In these coordinates, an ECM $E=A \exp (i\omega t +
i\psi)$ is transformed into a family of fixed points $E_0=A \exp (i\psi)$.
Splitting the complex electric field $E_{0j}=x_j+i\,y_j$ and using the vector
$\BX_j=\left(x_j, y_j, n_j\right)$ \Eqs(\ref{eq:laser}) can be
written in the form of \Eq(\ref{eq:general}) and the above analysis
applies. 
The eigenvalues of the Jacobian in the rotating frame then determine the ECM's transverse stability.
Figure \ref{fig:ecmdynamics}a displays the position of the ECMs in the $(\omega,
n)$-plane and their stability for a choice of parameters. The black trajectory
displays the projection of the symmetric variable $n_\BS$. 

The bubbling behavior in the CC regime and the correlation of the
desynchronization with the power dropouts in the LFF regime can be understood
as follows. In the CC regime the dynamics comprises chaotic itinerancy among
the modes and global antimode dynamics \cite{MUL98} (see
\Fig\ref{fig:ecmdynamics}a). The modes involved in the chaotic itinerancy are
transversally stable (blue circles). The antimodes on the other hand are
transversally unstable (red squares). Thus, when the trajectory approaches the
antimode, noise can lead to desynchronization and bubbling occurs. The yellow
diamonds in \Fig4a mark the onset of desynchronization, showing that bubbling
always occurs in the vicinity of the antimodes (independent of the power).
Please note that due to the role of noise not every approach to an antimode
results in a bubbling excursion. 

In the LFF regime \cite{SAN94} the dynamics is similar. The intensity buildup
process in between power dropouts is characterized by chaotic switching between
different attractor ruins (ghosts) of unstable ECMs with a drift towards the
ECM with minimal $n$. All ECMs involved in the buildup process are
transversally stable and we observe no desynchronization.  After a transient
time, a power dropout takes place. During the dropout the trajectory collides
with an antimode in a crisis. Again, the vicinity to transversally unstable
antimodes - rather than the drop in power - leads to bubbling behavior.

The transverse stability of the ECMs depends on the laser and coupling
parameters as well as on the parameters of the particular ECM. Note that modes
and antimodes are not necessarily transversally stable or unstable,
respectively. The modes on the lower right-hand side in
\Fig\ref{fig:ecmdynamics}a, for instance, are transversally unstable. With
decreasing coupling strength $K$, more modes become transversally unstable
until the whole chaotic attractor loses its transversal stability.  This leads
to the blowout bifurcation $B2$ in \Fig\ref{fig:lyap}.

With increasing feedback strength the bubbling occurs less frequently and the
average synchronization interval $\Delta$ increases; however, we did not find 
a transition to a bubbling-free state in a physically reasonable range of $K$.
Note that neither $K$ nor the other parameters of our model are \textit{normal} parameters
in the sense of Ref.~\cite{TER99}. Thus we do not observe power-law scaling of 
$\Delta$ as in \cite{SAU98,VEN96a}.
The parallel Lyapunov exponent
$\lambda_{||}$ approaches zero with increasing $K$ and the chaoticity
decreases, making this situation less interesting for chaos-based applications. 

If the elements are coupled via an active relay, the synchronized lasers behave
like two delay-coupled lasers (see \Eq(\ref{eq:DynActiveRelay})). If we choose
$\Bf = \Bg$ and $K=L$, we obtain a system of two identical mutually coupled
semiconductor lasers, which has been studied before \cite{MUL04,ERZ05,ERZ06a}.
Such a system has rotating wave solutions of the form $E_{\BS} (t) = A_{\BS}
\exp(i\omega t)$, $E_{\BY} (t) = A_{\BY} \exp(i\omega t +i \psi)$, $n_{\BS} (t)
= n_{\BS}$, $n_{\BY}(t)=n_{\BY}$, called compound laser modes (CLMs).  Their
spectrum is more complex than for the ECMs: besides the synchronized solutions
(which correspond to ECMs), there exist antisymmetric modes, for which the
relay and the synchronized solution are in anti-phase ($\psi=\pi$), as well as
asymmetric modes where the relay has a different intensity than the outer
lasers.

The positions of the transversally unstable modes are close to those of the
ECMs of a single laser in the $(\omega,n)$ parameter space. Also the dynamics
of three identical coupled lasers is similar to the behavior in the presence of
a passive relay.  Indeed, we find bubbling in both the LFF and CC regime.  

In the experiments reported in \cite{FIS06} all the coupling parameters in the
setup are chosen identical, \ie, $L=2K$ in \Eqs(\ref{eq:general}) and
(\ref{eq:active}). But also in this case we observe qualitatively similar laser
dynamics, with a trajectory in parameter space coming close to the
transversally unstable CLMs.

To suppress the bubbling while maintaining strong chaos, we apply a
sufficiently higher pump current to the relay laser ($p_{\mbox{\scriptsize
relay}}=4.0$) than to the outer lasers ($p=1.0$). For this configuration we
have calculated $\lambda_{||}\approx0.026$, $\lambda_\perp\approx-0.032$,
confirming that the system is in the chaotic regime (cf.~\Fig\ref{fig:lyap}b).
The system still itinerates among the compound laser modes, but there is no
global antimode dynamics.  Moreover, in contrast to the behavior for the
symmetric case $p_{\mbox{\scriptsize relay}}=1.0$, the active relay now
suppresses the bubbling and there is no desynchronization (see
\Fig\ref{fig:ecmdynamics}b).  Inspecting \Fig\ref{fig:ecmdynamics}b, we can
conclude that the CLMs involved in the dynamics are indeed transversally
stable. If the middle laser is pumped less strongly than the outer ones, the
opposite effect is observed.

In conclusion, we have demonstrated a mechanism for desynchronization by
bubbling in a very general setting of two delay-coupled lasers with either
passive or active relay.  We have shown that in the CC and LFF regimes the
occurrence of bubbling is related to the transverse instability of some of the
compound cavity's antimodes, and that, by tuning of the active relay, it is
possible to suppress the bubbling. These synchronization properties are
decisive for the setup of chaos-synchronization based applications and provide
a strategy how to achieve stable synchronization.

\begin{acknowledgments} We thank P.~Ashwin, T.~Gavrielides, and C.~Mirasso for
  fruitful discussions. O.D.~acknowledges the Research Foundation Flanders
  (FWO-Vlaanderen) for her fellowship and for project support.  This work was
  partially supported by the Belgian Science Policy Office under grant IAP-VI10
  "photonics@be", by the EC Project GABA FP6-NEST contract 043309, and by DFG
  in the framework of Sfb 555.  \end{acknowledgments}


\begin{thebibliography}{10}

\bibitem{BOC02}
S. Boccaletti, J. Kurths, G. Osipov, D.~L. Valladares, and C.~S. Zhou, Phys.
  Rep. {\bf 366},  1  (2002).

\bibitem{PIK01}
A. Pikovsky, M.~G. Rosenblum, and J. Kurths, {\em Synchronization, A Universal
  Concept in Nonlinear Sciences} (Cambridge University Press, Cambridge, 2001).

\bibitem{PEC90}
L.~M. Pecora and T.~L. Carroll, Phys.~Rev.~Lett. {\bf 64},  821  (1990).

\bibitem{ASH94}
P. Ashwin, J. Buescu, and I. Stewart, Phys.~Lett.~A {\bf 193},  126  (1994).

\bibitem{TER99}
J.~R. Terry, K.~S. Thornburg, D.~J. DeShazer, G.~D. VanWiggeren, S. Zhu, P.
  Ashwin, and R. Roy, Phys. Rev. E {\bf 59},  4036  (1999).

\bibitem{SAU98}
M. Sauer and F. Kaiser, Phys. Lett. A {\bf 243},  38  (1998).

\bibitem{GAU96}
D.~J. Gauthier and J.~C. Bienfang, Phys. Rev. Lett. {\bf 77},  1751  (1996).

\bibitem{MUL04}
J. Mulet, C.~R. Mirasso, T. Heil, and I. Fischer, J. Opt. B {\bf 6},  97
  (2004).

\bibitem{SHA06}
L.~B. Shaw, I.~B. Schwartz, E.~A. Rogers, and R. Roy, Chaos {\bf 16},  015111
  (2006).

\bibitem{KLE06}
E. Klein, N. Gross, M.~G. Rosenblum, W. Kinzel, L. Khaykovich, and I. Kanter,
  Phys.~Rev.~E {\bf 73},  066214  (2006).

\bibitem{LAN07}
A.~S. Landsman and I.~B. Schwartz, Phys.~Rev.~E {\bf 75},  026201  (2007).

\bibitem{FIS06}
I. Fischer, R. Vicente, J.~M. Buld{\'u}, M. Peil, C.~R. Mirasso, M.~C. Torrent,
  and J. Garc{\'i}a-Ojalvo, Phys.~Rev.~Lett. {\bf 97},  123902  (2006).

\bibitem{VIC04}
R. Vicente, C.~R. Mirasso, and I. Fischer, Opt. Lett. {\bf 32},  403  (2004).

\bibitem{AHL98}
V. Ahlers, U. Parlitz, and W. Lauterborn, Phys. Rev. E {\bf 58},  7208  (1998).

\bibitem{VEN96a}
S.~C. Venkataramani, B.~R. Hunt, E. Ott, D.~J. Gauthier, and J.~C. Bienfang,
  Phys. Rev. Lett. {\bf 77},  5361  (1996).

\bibitem{LAN80b}
R. Lang and K. Kobayashi, IEEE J. Quantum Electron. {\bf 16},  347  (1980).

\bibitem{ALS96}
P.~M. Alsing, V. Kovanis, A. Gavrielides, and T. Erneux, Phys. Rev. A {\bf 53},
   4429  (1996).

\bibitem{FAR82}
J.~D. Farmer, Physica~D {\bf 4},  366  (1982).

\bibitem{OTT94}
E. Ott and J.~C. Sommerer, Phys. Lett. A {\bf 188},  39  (1994).

\bibitem{MOR92}
J. M{\o}rk, B. Tromborg, and J. Mark, IEEE J. Quantum Electron. {\bf 28},  93
  (1992).

\bibitem{YAN05}
S. Yanchuk, Math.~Meth.~Appl.~Sci. {\bf 28},  363  (2005).

\bibitem{MUL98}
J. Mulet and C.~R. Mirasso, Phys. Rev. E {\bf 59},  5400  (1999).

\bibitem{SAN94}
T. Sano, Phys. Rev. A {\bf 50},  2719  (1994).

\bibitem{ERZ05}
H. Erzgr{\"a}ber, D. Lenstra, B. Krauskopf, E. Wille, M. Peil, I. Fischer, and
  W. Els{\"a}{\ss}er, Opt.~Commun. {\bf 255},  286  (2005).

\bibitem{ERZ06a}
H. Erzgr{\"a}ber, B. Krauskopf, and D. Lenstra, SIAM J.~Appl.~Dyn.~Syst. {\bf
  5},  30  (2006).

\end{thebibliography}

\end{document}